\documentclass[12pt]{iopart}

\expandafter\let\csname equation*\endcsname\relax
\expandafter\let\csname endequation*\endcsname\relax

\usepackage[font=small]{caption}
\usepackage{
,graphicx
,color
,latexsym
,mathtools
,subfig
}

\makeatletter

\newcommand{\Rmnum}[1]{\expandafter\@slowromancap\romannumeral #1@}
\makeatother

\let\origdescription\description
\renewenvironment{description}{
  \setlength{\leftmargini}{0em}
  \origdescription
  \setlength{\itemindent}{0em}
  \setlength{\labelsep}{\textwidth}
}
{\endlist}

\begin{document}

\title{Characterising Transient Noise in the LIGO Detectors}

\author{L. K. Nuttall$^{1}$ for the LIGO Scientific Collaboration}

\address{$^{1}$Cardiff University, Cardiff, CF24 3AA, United Kingdom}

\ead{laura.nuttall@ligo.org}

% Abstract text to be placed here %%%%%%%%%%%%
\begin{abstract}
Data from the LIGO detectors 
typically contain many non-Gaussian noise transients which 
arise due to instrumental and environmental conditions. These non-Gaussian 
transients can be an issue for the modelled and unmodelled transient 
gravitational-wave searches, as they can mask or mimic a true signal. Data 
quality can change quite rapidly, making it imperative to track and find new 
sources of transient noise so that data are minimally contaminated. Several 
examples of transient noise and the tools used to track them are presented. 
These instances serve to highlight the diverse range of noise sources 
present at the LIGO detectors during their second observing run.
\end{abstract}

\maketitle

\section{Introduction}
The Advanced Laser Interferometer Gravitational-wave Observatory (LIGO) 
detectors~\cite{Harry:lig} are some of the most sophisticated scientific 
instruments in the world. They are designed to measure stretching and 
squeezing of spacetime due to the passage of gravitational waves like those 
produced in 
the collisions of black holes and neutron stars. The two LIGO detectors are 
twin dual-recycled Michelson interferometers located in Hanford, WA and 
Livingston, LA~\cite{Aasi:ali}. The interferometers are L-shaped, with arms 
stretching 4 km in length; we often refer to each arm separately as the X- and 
Y-arm. Since they began operations in 2015 they have 
completed two observing runs. The first observing run (O1) spanned from 
September 2015 to January 2016, accumulating approximately 49 days of 
coincident data. The second observing run (O2) accumulated 117 days of 
coincident data, 
from November 2016 to August 2017. Advanced Virgo~\cite{Acernese:vir}, located 
in Italy, also joined the Advanced 
interferometric network for the final month of O2. To date, five
gravitational-wave signals and one candidate signal have been detected from 
the merger of black holes, with progenitor masses spanning 7-36~M$_{\odot}$~\cite{GW150914,GW151226,LVT151012,GW170104,GW170814, GW170608}. 
The first gravitational-wave signal from the coalescence of two neutron stars 
was 
recently announced~\cite{GW170817}, with much excitement for the coincident 
observations in the electromagnetic spectrum by many independent telescope 
facilities~\cite{MMA}.

A clear understanding of these gravitational-wave detectors is paramount to 
measuring such signals~\cite{GW150914_DQ}. 
The interferometers contain state-of-the-art hardware 
to isolate the instruments from their local environments and augment their 
interaction with a passing gravitational wave. The strong GW150914 signal only 
produced a relative length change, or strain, of the LIGO detector arms by 
1 part in 
$10^{21}$. Environmental and instrumental noise can cause relative 
length changes similar to or larger than true astrophysical signals. 
In addition, the characteristics of the detector output signal change on a 
daily basis due to environmental and hardware issues which unexpectedly 
arise. There are over two hundred thousand 
`auxiliary channels' which measure environmental and instrumental behaviour for 
each of the LIGO sites. These channels enable researchers to locate the source 
of noise so that detector sensitivity can be restored and the output data are 
minimally affected.

LIGO data are typically non-stationary and non-Gaussian; detector noise can 
have varying effects on different types of 
gravitational-wave searches. Long duration gravitational-wave searches, such 
as those searching for signals from pulsars or the stochastic background, are 
most susceptible to sources of noise which manifest themselves at a 
specific frequency or which produce combs in the amplitude spectral 
density~\cite{CW_detchar}. In contrast, searches for transient gravitational 
waves, both modelled and unmodelled\footnote{Modelled refers to searches for 
gravitational waves from compact binary coalescences with a known waveform 
morphology, whereas unmodelled refers to searches for gravitational-wave bursts of no specific waveform morphology.}, are most affected by non-Gaussian 
noise transients or `glitches'~\cite{GW150914_DQ,CBC_DQ}. 
This noise can mask or mimic a transient 
gravitational-wave signal. However with the correct understanding and treatment 
of noise, the sensitivity of gravitational-wave searches can be vastly 
improved. For example, one search for compact binary coalescences had as much 
as a 90\% improvement in its sensitivity\footnote{Search sensitivity refers to volume x time at a specific inverse false alarm rate whereas detector sensitivity refers to strain/$\sqrt{\mathrm{Hz}}$.} in parts of O1~\cite{CBC_DQ}. 

This paper aims to highlight examples of non-Gaussian noise transients seen 
in the LIGO detectors which have adversely affected the search for transient 
gravitational waves in O2. Complementary examples are also presented 
in~\cite{Beverly_DQ}.

\section{Example Tools used to Characterise the LIGO Data}
\begin{figure}[!h]
\centering
\includegraphics[width=2.6in]{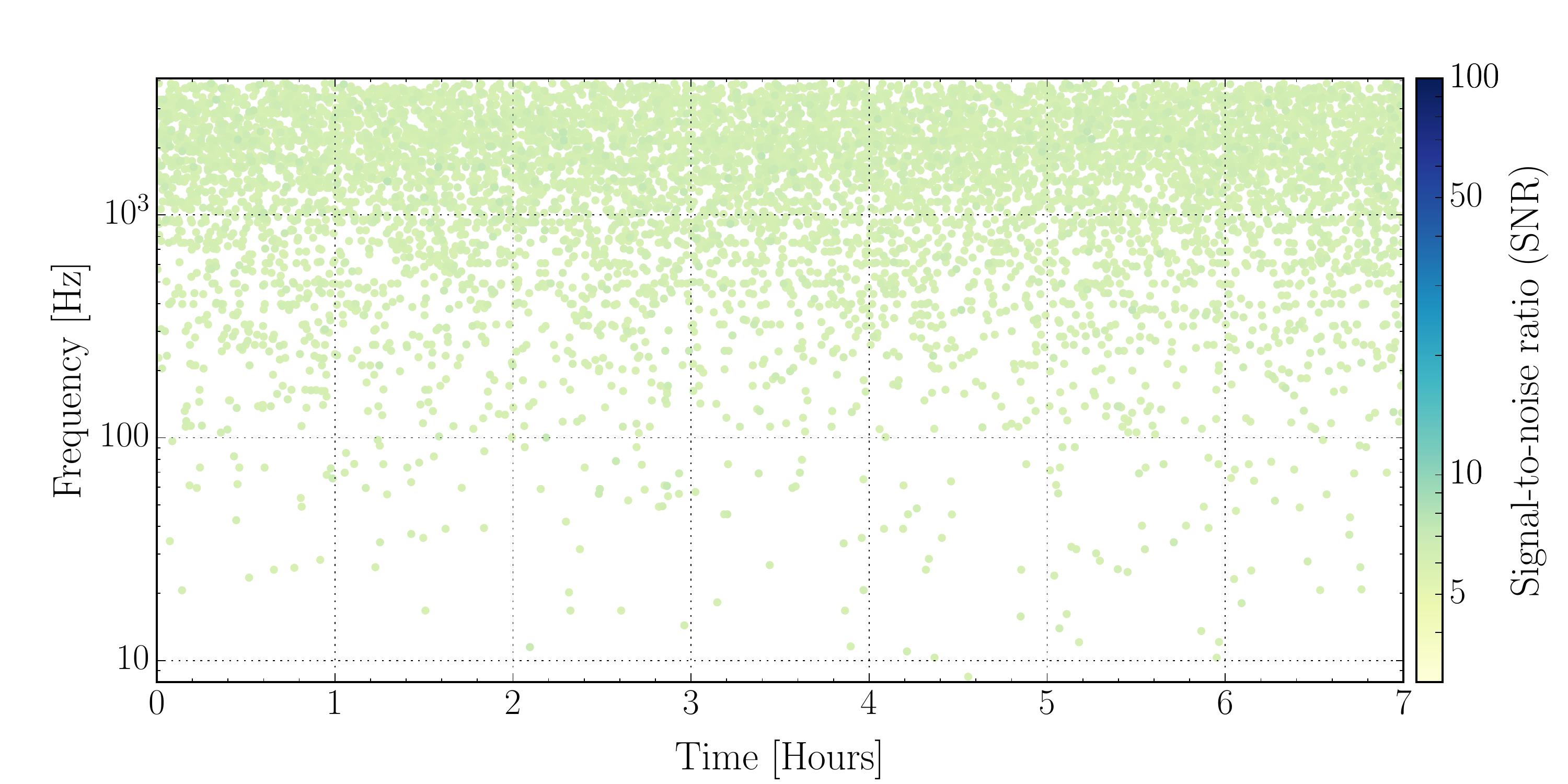}
\includegraphics[width=2.6in]{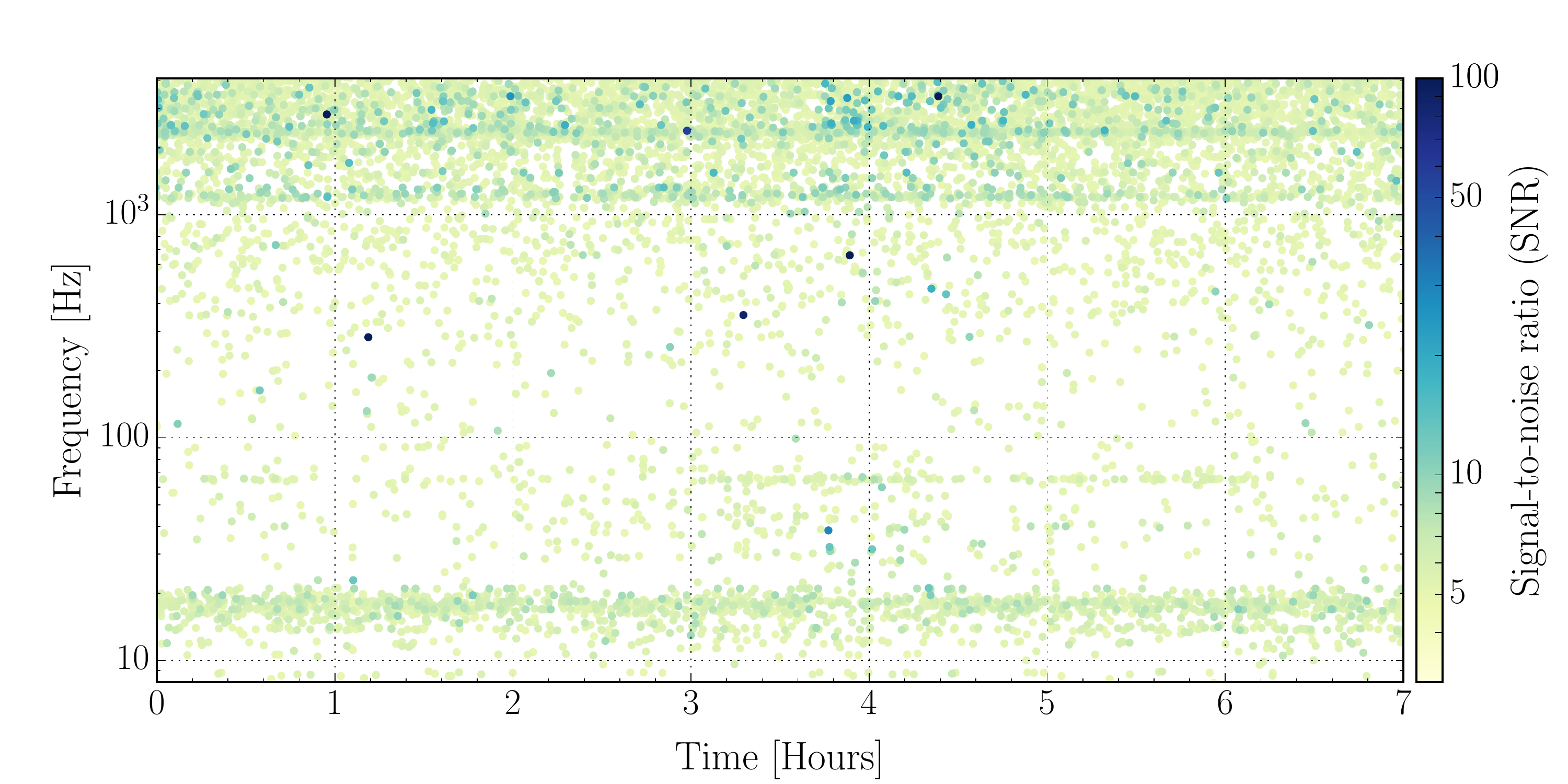}
%%% where xxxxxx name represents "figurename.eps"
\caption{The distribution of Omicron triggers in frequency and SNR over a seven 
hour time period on Gaussian (left) and typical O2 (right) LIGO data. Note the 
log scale makes Omicron triggers appear more dense at high frequencies.}
\label{fig:data}
\end{figure}

A typical method to visualise the LIGO data from a transient search 
perspective is by representing each detector's data by an unmodelled or 
`burst' transient identification 
algorithm. A method commonly used to characterise LIGO data is Omicron, which 
identifies excess power 
transients using a generic sine Gaussian time-frequency 
projection and reports transients at a given central time, central frequency, 
duration, bandwidth, Q-value and signal-to-noise ratio (SNR) 
(or normalised tile energy)~\cite{Omicron}. 
Figure~\ref{fig:data} shows Omicron applied to LIGO data 
in the ideal case of Gaussian noise (left) and typical operations (right). The 
Gaussian noise plot shows few features across the time-frequency space, 
whereas the typical data plot shows much more structure due to various 
issues taking place throughout the day. Examples of the nature of these 
glitches will be presented in the next Section. 

The main method to find the cause of transient noise in the gravitational-wave 
channel is to find time coincidences with auxiliary channels. For example, the 
Hierarchical Veto (HVeto) algorithm looks for time coincidences between 
glitches in auxiliary channels and the gravitational-wave channel~\cite{Hveto}. This method 
firstly uses Omicron to identify transients in auxiliary channels before 
time-correlating them with the gravitational-wave channel. Many examples 
presented in the next section made use of HVeto to identify a suspect area of 
the detector to investigate.

LigoDV-web is a software suite that allows easy access to 
data~\cite{ligodvweb}. Through its web interface, grabbing and 
viewing data is exceptionally simple; LigoDV-web is often initially used to 
view a variety of data, whether from the gravitational-wave or auxiliary 
channels. 

GravitySpy is a citizen science project, aiming to classify glitches 
within the LIGO data using human classification to train a convolutional 
neural network~\cite{gravityspy}. 
This new tool has proved 
extremely useful in charactering glitches identified by Omicron, and grouping 
them in to known classes. A database now exists of all the glitches in the 
first and second observing runs. Investigations utilising this database 
are currently underway.

Many of the tools used to characterise the LIGO data are collated on the 
internal LIGO summary pages. A public viewable version 
of these pages is available at~\cite{Public_summary}. These pages house key 
information in an easily viewable format to keep track of the varying data 
quality of the detectors. The summary pages are vital in quickly identifying 
and finding the source of new noise. These pages and the figures presented in 
this paper make use of the GWpy software package~\cite{GWpy}. 

\section{Transient Noise Examples in LIGO}

\subsection{Thirsty Ravens}
In the summer of 2017 many glitches around 90~Hz 
in the gravitational-wave channel 
were correlated with a microphone at the Y-end of the Hanford interferometer 
by HVeto. Listening to the microphone output at the time of these 
glitches implicated ravens that had been seen outside the Y-end of the 
detector. At the Y-end are cryopumps which store nitrogen for various 
detector operations. Ice accumulates on the vent lines transporting the 
nitrogen in to the Y-end building. Upon inspection of these vent lines, peck 
marks were observed consistent with the size of a raven's beak. To test this 
hypothesis imitation pecking was simulated by chipping at the ice and was 
found to be consistent with the observed noise. 
Further investigations concluded that the 
ravens peck the vent line which is connected to and vibrates the vacuum 
enclosure and other instrumental components. This causes the optical path 
length of the main laser beam to vary. The laser light is scattered from the 
Y-end test mass, is reflected off other surfaces and recombines in to the main 
laser beam causing noise pollution~\cite{Ravens:alog}. 
The source of this noise can be easily 
solved by insulating the vent lines to prevent ice accumulating or adding an 
additional casing around the vent lines to prevent the ravens accessing the 
ice. There are plans to implement one or both of the proposed mitigation 
solutions before the next observing run 
(expected late 2018)~\cite{living_review}.

\subsection{Optical Lever Glitches}
Optical lever lasers are used to sense and stabilise the angular alignment 
of interferometer optics. In February 2017, HVeto 
discovered time coincidences between glitches seen in the Hanford 
gravitational-wave channel and a channel which monitors the power to the 
optical lever at the Y-end test mass. 
These glitches were found to adversely affect the modelled transient 
searches, as the transient noise manifested itself across frequencies 
< 200~Hz and were sufficiently loud (i.e. a high SNR). 
It was found that the optical lever laser was glitching, and the 
radiation pressure from this glitching was coupling in to the 
gravitational-wave channel. The output power of the laser was adjusted to 
return to an almost glitch-free operating power. 
At various points in the second 
observing run the optical lever laser at the intermediate Y-test mass also 
needed power adjustments and the X-end laser needed replacing. 

\subsection{Magnetic Glitches}
\begin{figure}[!h]
\centering
\includegraphics[width=5.5in]{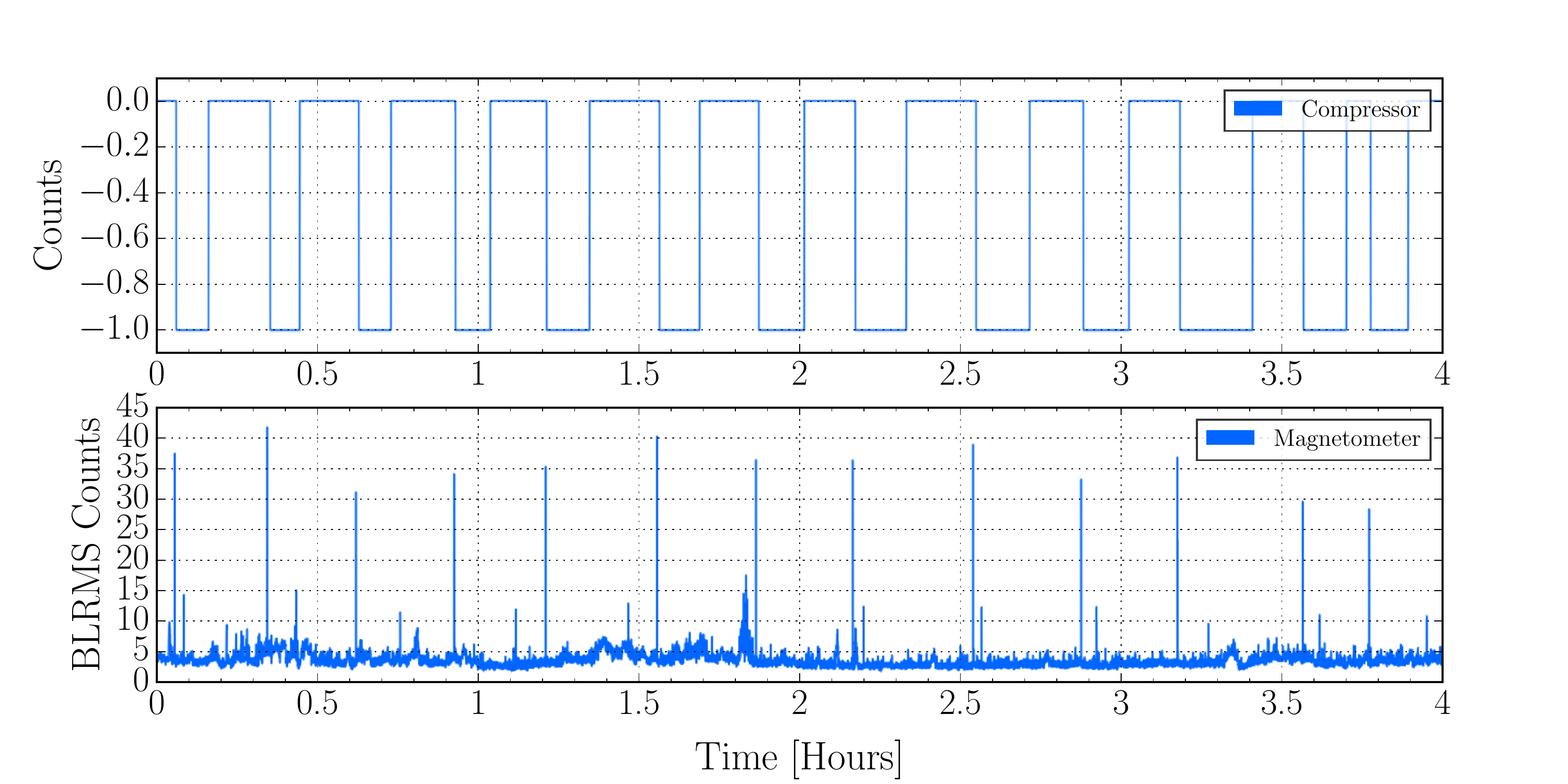}
\caption{The time coincidence between the switching on of a compressor and 
glitches in a magnetometer at the X-end. Top: Time series of a channel 
which monitors the state of the compressor. Bottom: The band limited 
root-mean-square (BLRMS) series over a channel which monitors an X-end 
magnetometer. The BLRMS between 75-100~Hz calculated once per second of this 
magnetometer channel provided good evidence for the link between the 
switching on of the compressor and glitches in 
the magnetometer and, subsequently, the gravitational-wave channel.}
\label{fig:magglitches}
\end{figure}

In February 2017 a group of glitches around 50-60~Hz appeared at the 
Livingston detector which accounted for approximately 15\% of all glitches 
between 10-1000~Hz. These correlated in time with glitches in 
magnetometers and channels monitoring the AC 
power mains at the X-end; they were 
initially identified by HVeto~\cite{Chiller:alog}. Further investigations 
found that the glitches related to the switching on of a 
compressor, as shown in Figure~\ref{fig:magglitches}. To mitigate 
the effect of these glitches grounding modifications to the X-end 
electronics rack were needed. Since this 
work, the 50-60~Hz glitches disappeared and have not been seen since at 
Livingston.

\subsection{Scattered Light}
\begin{figure}[!h]
\centering
\includegraphics[width=2in]{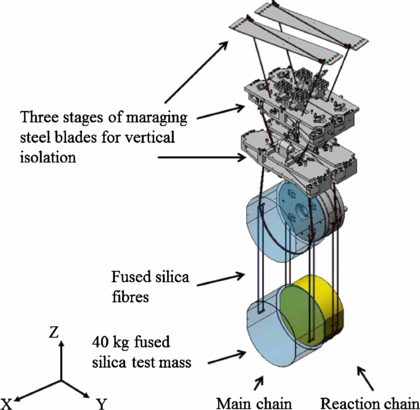}
\includegraphics[width=3in]{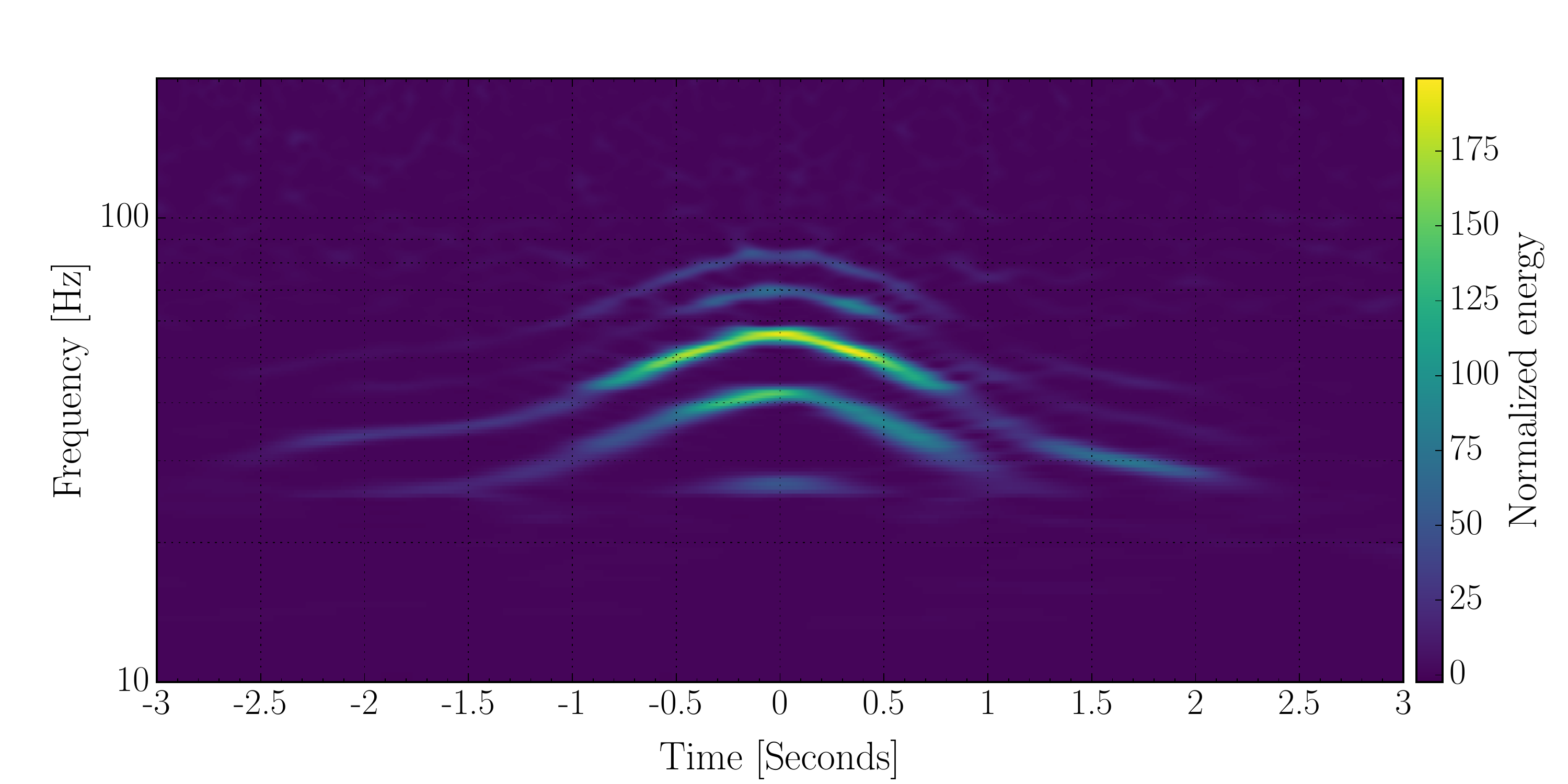}
\caption{Left: Schematic of the quadruple pendulum and quadruple reaction 
pendulum~\cite{suspension_paper}. The reaction mass is coloured in yellow. 
Right: Normalised spectrogram of 
gravitational-wave data with scattering noise present. 
Characteristics of scattering noise are arches. This figure shows arches 
affecting several seconds of data between 20-90~Hz.}
\label{fig:scatter}
\end{figure}

Light scattering is a common source of noise at both interferometers. 
Any motion 
or slight misalignment can cause parts of the main laser beam to scatter off 
moving surfaces, in any of the several chambers, and couple back in to the 
main beam. This example concerns the Y-end reaction mass at Livingston, which 
on occasions prior to February 2017, caused scattering in the 
gravitational-wave channel up to $\sim$90~Hz. Figure~\ref{fig:scatter} (left) 
shows the location of the reaction mass, in yellow, in relation to the main 
test mass and quadruple pendulum. 
The scattering arches seen in the gravitational-wave channel are shown 
in the right of 
Figure~\ref{fig:scatter}. These are very detrimental to the 
transient searches, both modelled and unmodelled, which typically analyse data 
above 20~Hz. The LIGO summary pages have a monitor for tracking scattering 
from a number of optics throughout each detector. This monitor searches for 
evidence of beam scattering based on the velocity of an optic's motion, and 
the scattering fringe frequency is predicted from~\cite{virgo:scatter}. This 
monitor originally suggested an optic at the Y-end as the source of the scattering. After further investigation the reaction mass, 
which serves to control the position of the Y-end test mass, was implicated. 
By changing the angle of the reaction mass, the scattering issue was 
resolved~\cite{scatter:alog}. During upgrades, before the start of the third 
observing run, many baffles 
throughout each detector are being installed to mitigate the effects of 
scattering~\cite{Aasi:ali}.

\subsection{Thunderstorms}
Throughout the second observing run thunderstorms near the Livingston 
detector have had a negative impact on sensitivity. One way in which 
sensitivity is measured is by monitoring the binary neutron star inspiral 
range; the range at which a detection of a 1.4-1.4 M$_\odot$ neutron-star 
merger is made at a 
SNR of 8. Thunderstorms have resulted in a range decrease of 10~Mpc, 
approximately a 10\% decrease, by causing an increase in noise between 
40-60~Hz. This is caused by acoustic noise coupling in to the detector, causing 
vibrations and thus light scattering noise. 
Upon investigating the timing of the thunder claps in various microphones 
around the detector, it was found that acoustic noise 
coupled to a baffle between the two chambers which house a three-optic cavity 
that cleans the input laser light before it enters the interferometer 
arms~\cite{Aasi:ali, swiss:lloalog}. This baffle was also a dominant source 
of vibrational noise at Hanford~\cite{swiss:lhoalog}. 
Raising the resonant frequency of this 
baffle will reduce the scattering effect. During upgrades to the detectors 
before the start of the third observing run, the shiny parts of this baffle, 
which cause the most scattering, will be removed.

\subsection{Excess Ground Motion}
Throughout the second observing run excess ground motion at both sites has 
been shown to have an adverse affect on detector sensitivity. 
Different types of 
activity around each site can have varying effects, although ground motion 
due to earthquakes are a constant challenge. In addition the Hanford 
detector can experience winds with speeds in excess of 50 mph. 
During the winter months,
which brought snow to the Hanford site, detector sensitivity was particularly 
impacted due to plows clearing the snow. This activity caused excess 
ground motion between 10-30~Hz measured by a seismometer located at the 
building which 
house the main detector optics. This noise could then be seen coupling in to 
the gravitational-wave channel, thereby affecting the inspiral range. These 
times of excess noise can often be detrimental to the transient searches, and 
motivating the removal of times associated with excess ground motion during 
O2. We expect ground motion to continue to adversely affect the astrophysical 
searches during future observing runs.

\section{Concluding remarks}
This paper presents some examples of transient noise seen in the LIGO 
detector during O2. 
In some instances the sources of noise could be fixed at the 
source, whereas others could simply be further understood and planned to be 
ameliorated in upcoming upgrade work, currently taking place at both sites. 
Another mitigation strategy to already collected data, is the use of data 
quality vetoes to remove or down-rank egregious data used in transient 
searches. Examples of this strategy are presented in~\cite{GW150914_DQ, CBC_DQ}.

As the LIGO detectors are upgraded and commissioned to design sensitivity, new 
sources of noise will be uncovered. See for instance the different noise 
examples presented as the detectors were being commissioned before 
O1~\cite{Nuttall:er6}, during O1\cite{GW150914_DQ,CBC_DQ} and during O2 (this 
paper). At the same time, the rate of detectable 
gravitational-wave signals will also increase. Therefore the chances they will 
overlap with transient noise also increase, particularly for longer duration 
(minute time-scale) signals. It is extremely important to investigate ways 
of tracking new groups of transient noise and methods for excising them from 
future data so the astrophysical potential of the LIGO detectors is realised.

\section{Acknowledgments}
The author thanks Thomas Massinger, Jess McIver and Duncan Macleod for feedback on this paper, and the LIGO Detector Characterisation Group for many fruitful discussions. LKN received funding from the European Union Horizon 2020 research and innovation programme under the Marie Skłodowska-Curie grant agreement No 663830.\\

%%%%%%%%%% Insert bibliography here %%%%%%%%%%%%%%


\begin{thebibliography}{9}

\bibitem{Harry:lig} Harry, G. M. for The LIGO Scientific Collaboration. 2010. Advanced LIGO: The Next Generation of Gravitational Wave Detectors. Class. Quantum Grav. 27, 084006
\bibitem{Aasi:ali} Aasi, J. et al. 2015. Advanced LIGO. Class.Quant.Grav. 32, 074001
\bibitem{Acernese:vir} Acernese, F. et al. 2015. Advanced Virgo: A 2nd Generation Interferometric Gravitational Wave Detector, Class. Quantum Grav. 32, 024001
\bibitem{GW150914} Abbott, B. P. at al. 2016. Observation of Gravitational Waves from a Binary Black Hole Merger. Phys. Rev. Lett. 116, 061102
\bibitem{GW151226} Abbott, B. P. at al. 2016. GW151226: Observation of Gravitational Waves from a 22-Solar-Mass Binary Black Hole Coalescence, Phys. Rev. Lett., 116, 241103
\bibitem{LVT151012} Abbott, B. P. at al. 2016. Binary Black Hole Mergers in the First Advanced LIGO Observing Run, Phys. Rev. X., 6, 041015
\bibitem{GW170104} Abbott, B. P. et al. 2017. GW170104: Observation of a 50-Solar-Mass Binary Black Hole Coalescence at Redshift 0.2. Phys. Rev. Lett. 118, 221101
\bibitem{GW170814} Abbott, B. P. et al. 2017. GW170814: A Three-Detector Observation of Gravitational Waves from a Binary Black Hole Coalescence. Phys. Rev. Lett. 119. 141101
\bibitem{GW170608} Abbott B. P. et al. 2017, GW170608: Observation of a 19-solar-mass Binary Black Hole Coalescence. Astrophys. J. 852, L35
\bibitem{GW170817} Abbott, B. P. et al. 2017. GW170817: Observation of Gravitational Waves from a Binary Neutron Star Inspiral. Phys. Rev. Lett. 119. 161101
\bibitem{MMA} Abbott, B. P. et al. 2017. Multi-messenger Observations of a Binary Neutron Star Merger. ApJL, 848, L2
\bibitem{GW150914_DQ} Abbott, B. P. 2016. Characterization of Transient Noise in Advanced LIGO relevant to Gravitational Wave Signal GW150914, Class. Quantum Grav., 33, 134001
\bibitem{CW_detchar} Aasi, J. et al. 2015. Characterization of the LIGO Detectors during their Sixth Science Run. Class. Quant. Grav. 32 115012
\bibitem{CBC_DQ} Abbott, B. P. 2018. Effects of Data Quality Vetoes on a Search for Compact Binary Coalescences in Advanced LIGO's First Observing Run, Class. Quant. Grav. 35  065010
\bibitem{Beverly_DQ} Berger, B. K. 2018. Identification and Mitigation of Advanced LIGO Noise Sources. J. Phys.: Conf. Ser. 957 012004
\bibitem{Omicron} Robinet, F. 2015. Omicron: An Algorithm to Detect and Characterize Transient Noise in Gravitational-Wave Detectors. https://tds.ego-gw.it/ql/?c=10651
\bibitem{Hveto} Smith, J.R. et al. 2011. A Hierarchical Method for Vetoing Noise Transients in Gravitational-Wave Detectors. Class. Quantum Grav. 28 235005
\bibitem{ligodvweb} Areeda, J. S. et al. 2017. LigoDV-web: Providing Easy, Secure and Universal Access to a Large Distributed Scientific Data Store for the LIGO Scientific Collaboration. Astron. Comput. 18. 27-34
\bibitem{gravityspy} Zevin, M. et al. 2017. Gravity Spy: Integrating Advanced LIGO Detector Characterization, Machine Learning, and Citizen Science. Class. Quant. Grav. 34. 064003
\bibitem{Public_summary} LIGO Scientific Collaboration, https://losc-dev.ligo.org/detector\_status/. Accessed October 2017
\bibitem{GWpy} Macleod, D. M. https://gwpy.github.io/docs/stable/. Accessed October 2017
\bibitem{Ravens:alog} LIGO Scientific Collaboration, https://alog.ligo-wa.caltech.edu/aLOG/index.php?callRep=37630. Accessed October 2017
\bibitem{living_review} Abbott, B. P. 2016, Prospects for Observing and Localizing Gravitational-Wave Transients with Advanced LIGO, Advanced Virgo and KAGRA. Living Rev. Relativity 19, 1
\bibitem{Chiller:alog} LIGO Scientific Collaboration, https://alog.ligo-la.caltech.edu/aLOG/index.php?callRep=32389. Accessed October 2017
\bibitem{virgo:scatter} Accadia, T. et al. 2010. Noise from Scattered Light in Virgo's Second Science Run Data. Class. Quantum Grav. 27 194011
\bibitem{scatter:alog} LIGO Scientific Collaboration, https://alog.ligo-la.caltech.edu/aLOG/index.php?callRep=31597. Accessed October 2017
\bibitem{suspension_paper} Aston, S. M. et al. 2012, Update on Quadruple Suspension Design for Advanced LIGO. Class. Quantum Grav. 29. 235004
\bibitem{swiss:lloalog} LIGO Scientific Collaboration, https://alog.ligo-la.caltech.edu/aLOG/index.php?callRep=34161. Accessed October 2017
\bibitem{swiss:lhoalog} LIGO Scientific Collaboration, https://alog.ligo-wa.caltech.edu/aLOG/index.php?callRep=35735. Accessed October 2017
\bibitem{Nuttall:er6} Nuttall, L. K. et al. 2015, Improving the Data Quality of Advanced LIGO Based on Early Engineering Run Results. Class. Quantum Grav. 32. 245005

\end{thebibliography}
\end{document}